\documentclass[aps,prd,nofootinbib,twocolumn]{revtex4}
\usepackage{amssymb}
\usepackage{amsmath,color}
\usepackage{epsfig}
\usepackage{graphicx,epsf,epsfig}
\usepackage{bbm}
\usepackage{subfigure}
\usepackage{multirow}
\usepackage{setspace}
\usepackage{verbatim}
\usepackage{epstopdf}
\usepackage{ulem}
\usepackage[linktocpage]{hyperref}
\usepackage{multirow}
\usepackage[marginal]{footmisc}
\newcommand{\be}{\begin{eqnarray}}
\newcommand{\ee}{\end{eqnarray}}
\newcommand{\bq}{\begin{equation}}
\newcommand{\eq}{\end{equation}}
\newcommand{\bp}{\begin{split}}
\newcommand{\ep}{\end{split}}

\begin{document}
\title{An analytical approximation of the scalar spectrum in the ultra-slow-roll inflationary models}

\author{Jing Liu$^{1,2}$}
\email{liujing@itp.ac.cn}

\author{Zong-Kuan Guo$^{1,2,3}$}
\email{guozk@itp.ac.cn}

\author{Rong-Gen Cai$^{1,2,3}$}
\email{cairg@itp.ac.cn}

\affiliation{$^1$CAS Key Laboratory of Theoretical Physics, Institute of Theoretical Physics,
    Chinese Academy of Sciences, P.O. Box 2735, Beijing 100190, China}
\affiliation{$^2$School of Physical Sciences, University of Chinese Academy of Sciences,
    No.19A Yuquan Road, Beijing 100049, China}
\affiliation{$^3$ School of Fundamental Physics and Mathematical Sciences, Hangzhou Institute for Advanced Study, University of Chinese Academy of Sciences, Hangzhou 310024, China}
\begin{abstract}
The ultra-slow-roll~(USR) inflationary models predict large-amplitude scalar perturbations at small scales which can lead to the primordial black hole production and scalar-induced gravitational waves. In general scalar perturbations in the USR models can only be obtained using numerical method because the usual slow-roll approximation breaks.
In this work, we propose an analytical approach to estimate the scalar spectrum which is consistent with the numerical result. We find that the USR inflationary models predict a peak with power-law slopes in the scalar spectrum and energy spectrum of gravitational waves, and we derive the expression of the spectral indexes in terms of the inflationary potential. In turn, the inflationary potential near the USR regime can be reconstructed from the negative spectral index of the gravitational wave energy spectrum.

\end{abstract}

\maketitle

\section{Introduction}
\label{sec:int}

Primordial scalar perturbations from quantum fluctuations during inflation can explain the cosmic microwave background~(CMB) radiation anisotropy and seed the large-scale structure of the Universe observed today~\cite{Lewis:1999bs,Bernardeau:2001qr}.
The amplitude of the power spectrum of scalar perturbations is tightly constrained to be around $2.2\times10^{-9}$ by observations at large scales. However, at small scales the constraints on the scalar perturbations are very loose~\cite{Mesinger:2005ah,Bringmann:2011ut,Chluba:2012we}. The USR inflationary models predict a large-amplitude scalar power spectrum at small scales, which can result in interesting consequences. Overdense regions caused by large-amplitude scalar perturbations, when they reenter the horizon,  collapse into primordial black holes~(PBHs) if the average energy density of the region is above a certain threshold~\cite{Zeldovich:1967,Carr:1974nx,Hawking:1971ei}. Such PBHs can constitute dark matter without new physics or explain the merger events observed by LIGO~\cite{Green:2004wb,Frampton:2009nx,Carr:2009jm,Carr:2016drx,Gao:2018pvq,Nakama:2019htb,Fu:2019ttf,Bird:2016dcv,Sasaki:2016jop,Mishra:2019pzq,Cai:2019bmk}. Moreover, since scalar perturbations are coupled with tensor perturbations at the nonlinear order, amplified scalar perturbations can also result in induced gravitational waves~(GWs). Because of the weak coupling between tensor perturbations and the matter fields, GWs can penetrate through radiation without attenuation and thus carry information in the very early Universe. The frequency of the induced GWs can lie in the sensitive bounds of space-based and ground-based GW detectors, which open a window to detect scalar perturbations at small scales and give constraints on PBHs from the stochastic gravitational wave background~(SGWB)~\cite{Cai:2018dig,Bartolo:2018rku,Saito:2008jc,Saito:2009jt,Wang:2019kaf,Ananda:2006af,Baumann:2007zm,Cai:2019elf,Yuan:2019udt,Lu:2019sti}.

The USR models are realized in various scenarios, including string theory~\cite{Cicoli:2018asa}, supergravity~\cite{Gao:2018pvq,Addazi:2018pbg}, nonminimal derivative coupling models~\cite{Fu:2019vqc}, Higgs-$R^{2}$ inflation~\cite{Cheong:2019vzl}, critical Higgs inflation~\cite{Drees:2019xpp}, k/G inflation~\cite{Lin:2020goi}, $\alpha$-attractor models~\cite{Dalianis:2018frf}, non-minimal coupling $R^{2}$ gravity~\cite{Pi:2017gih} and inflection-point inflation~\cite{Ballesteros:2017fsr,Gong:2017qlj}. The inflationary potential is extremely flat in the USR regime. The slow-roll conditions are violated, so the modes of scalar perturbations continue evolving after they leave the Hubble horizon. In the USR models, numerical simulations are required to obtain the scalar spectrum because large discrepancies exist between the numerical result and the approximate result under the usual slow-roll approximations. In this paper, we consider an analytical method to obtain approximate result of the scalar spectrum, which is consistent with the numerical result, and obtain the scalar spectral index in terms of the inflationary potential. Based on the relationship between the spectral index of the scalar spectrum and energy spectrum of GWs claimed in Ref.~\cite{Xu:2019bdp}, we can obtain the spectral index of GW energy density from the inflationary potential and constrain the inflationary models from GWs. 


The paper is organized as follows. In Sec.~\ref{sec:sem}, we study the generation of scalar power spectrum from quantum fluctuations in the USR regime and analyse the superhorizon evolution of scalar perturbations. In Sec.~\ref{sec:PR}, using the results presented in Sec.~\ref{sec:sem}, we derive the explicit expression of spectral indexes of $\mathcal{P_{R}}$ in terms of the inflationary potentials. In Sec.~\ref{sec:GW}, using the formalism of calculating the energy spectrum of induced GWs in Ref.~\cite{Kohri:2018awv}, we give the numerical result of $\Omega_{\mathrm{GW}}$ in the cases of the numerical and approximate $\mathcal{P_{R}}$. In Sec.~\ref{sec:con}, we summarize our results.
We set $c=\hbar=8\pi G = 1$ throughout the paper.

\section{Evolution of scalar perturbations}
\label{sec:sem}

In this section, we briefly review the production of primordial scalar perturbations from quantum fluctuations during inflation, then find the approximate solution of the equation of motion~(EOM) of scalar perturbations in the USR inflationary models.

Now consider a single-field inflationary model in which the inflaton is minimally coupled to gravity. The effective action is given by
\begin{equation}
S=\int d^{4} x \sqrt{-g}\left[-\frac{R}{2} +\frac{1}{2} \partial_{\mu} \phi \partial^{\mu} \phi+V(\phi)\right],
\end{equation}
where $\phi$ denotes the inflaton field. In a Friedmann-Robertson-Walker Universe, the Friedmann equation and the EOM of $\phi$ read
\begin{equation}
\begin{split}
H^{2}=\dfrac{1}{3}\left(\dot{\phi}^{2}+V(\phi)\right),\\
\ddot{\phi}+3H\dot{\phi}+\dfrac{dV}{d\phi}=0,
\label{eq:eomphi}
\end{split}
\end{equation}
where dots denote the derivatives with respect to the cosmic time, and $H$ is the Hubble parameter.
The perturbed metric in the conformal Newtonian gauge can be expressed as
\begin{equation}
\label{eq:meper}
\begin{aligned}
d s^{2}=a^{2}(\tau)&\bigg\{-(1+2 \Phi) d \tau^{2}  \\
&\left.+\left[(1-2 \Phi) \delta_{i j}+\frac{1}{2} h_{i j}\right] d x^{i} d x^{j}\right\},
\end{aligned}
\end{equation}
where $\tau$ is the conformal time. Here we have neglected vector perturbations and the anisotropic stress. The gauge-invariant conformal curvature perturbation is
\begin{equation}
\mathcal{R}=-\Phi-\dfrac{\phi '}{\mathcal{H}}\delta\phi,
\end{equation}
where $\mathcal{H}\equiv aH$, $\delta\phi$ denotes the perturbation of $\phi$, a prime represents the derivative with respect to the conformal time $\tau$.
The action of scalar linear perturbations is~\cite{Kodama:1985bj,Mukhanov:1990me}
\begin{equation}
S=\frac{1}{2} \int\left[\left(u_{k}^{\prime}\right)^{2}+k^{2}u_{k}^{2}+\frac{z^{\prime \prime}}{z} u_{k}^{2}\right] d \tau d^{3} k,
\end{equation}
where $z\equiv \frac{a\dot{\phi}}{H}$, $u_{k}\equiv z\mathcal{R}_{k}$ and $\mathcal{R}=\int \frac{d^{3}k}{(2\pi)^{3/2}}\mathcal{R}_{k}e^{i\mathbf{k}\cdot \mathbf{x}}$. 
The EOM of $u_{k}$ is
\begin{equation}
u_{k}^{\prime \prime}+\left(k^{2}-\frac{z^{\prime \prime}}{z}\right) u_{k}=0,
\end{equation}
which is known as the Mukhanov-Sasaki equation. Equivalently,
\begin{equation}
\label{eq:tauRk}
\mathcal{R}_{k}^{\prime \prime}+2 \frac{z^{\prime}}{z} \mathcal{R}_{k}^{\prime}+k^{2} \mathcal{R}_{k}=0.
\end{equation}
The terms $\frac{z'}{z}$ and $\frac{z''}{z}$ can be expressed in terms of the slow-roll parameters
\begin{equation}
\begin{split}
\dfrac{z'}{z}&=\mathcal{H}(1+\epsilon-\eta),\\
\dfrac{z^{\prime \prime}}{z}&=2 \mathcal{H}^{2}\left(1-\frac{3}{2} \eta+\epsilon+\frac{1}{2} \eta^{2}-\frac{1}{2} \epsilon \eta+\frac{1}{2} \frac{1}{\mathcal{H}} \epsilon'-\frac{1}{2} \frac{1}{\mathcal{H}} \eta'\right),
\end{split}
\end{equation}
where $\epsilon$ and $\eta$ are defined by
\begin{equation}
\label{eq:srp}
\begin{split}
\epsilon&=-\dfrac{\dot{H}}{H^{2}}=\dfrac{3\dot{\phi}^{2}}{2V(\phi)+\dot{\phi}^{2}},\\
\eta&=-\dfrac{\ddot{\phi}}{\dot{\phi}H},
\end{split}
\end{equation}
and they are related by
\begin{equation}
\label{eq:rela}
\dot{\epsilon}=2 H \epsilon(\epsilon-\eta).
\end{equation}

Inflation is required to last for a sufficient long time to solve the horizon problem and flatness problem, which means $\epsilon$ must be small during inflation. 
Under the slow-roll conditions, $\epsilon,\eta\ll 1$. In the USR inflationary models,  $\eta$ changes quickly at the beginning and end of the USR regime, and remains almost constant during the USR regime. So we make the following assumptions. The USR region starts at $t_{s}$ and ends at $t_{e}$. During the USR regime, $\eta=\eta_{m}$, where $\eta_{m}$ is constant. The slow-roll conditions are valid outside the USR regime. 
In the following, we use $k_{s}$ and $k_{e}$ to denote the modes which cross the horizon at $t_{s}$ and $t_{e}$, respectively.

\begin{figure*}[t]
    \includegraphics[width=0.4\textwidth]{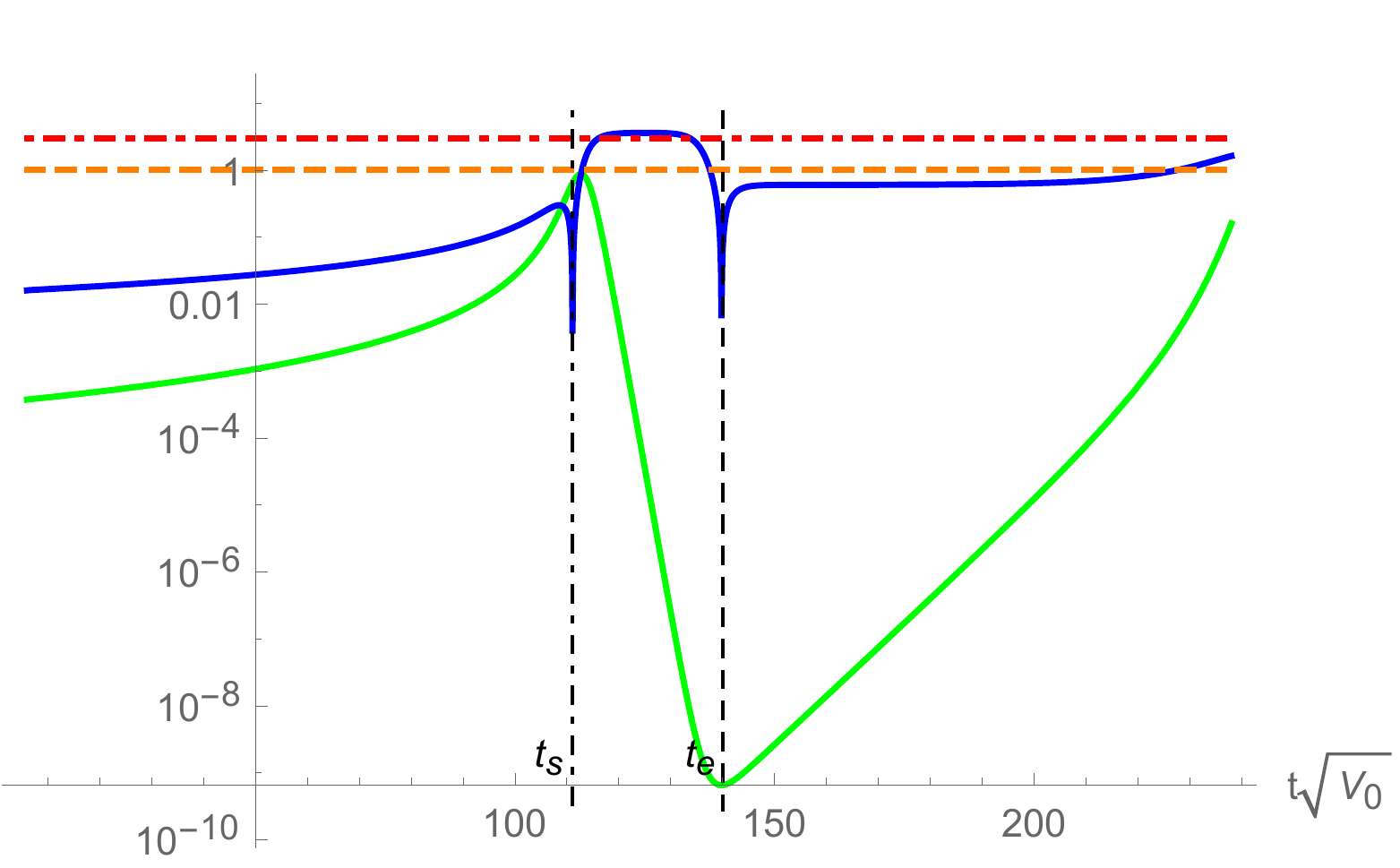}
    \includegraphics[width=0.38\textwidth]{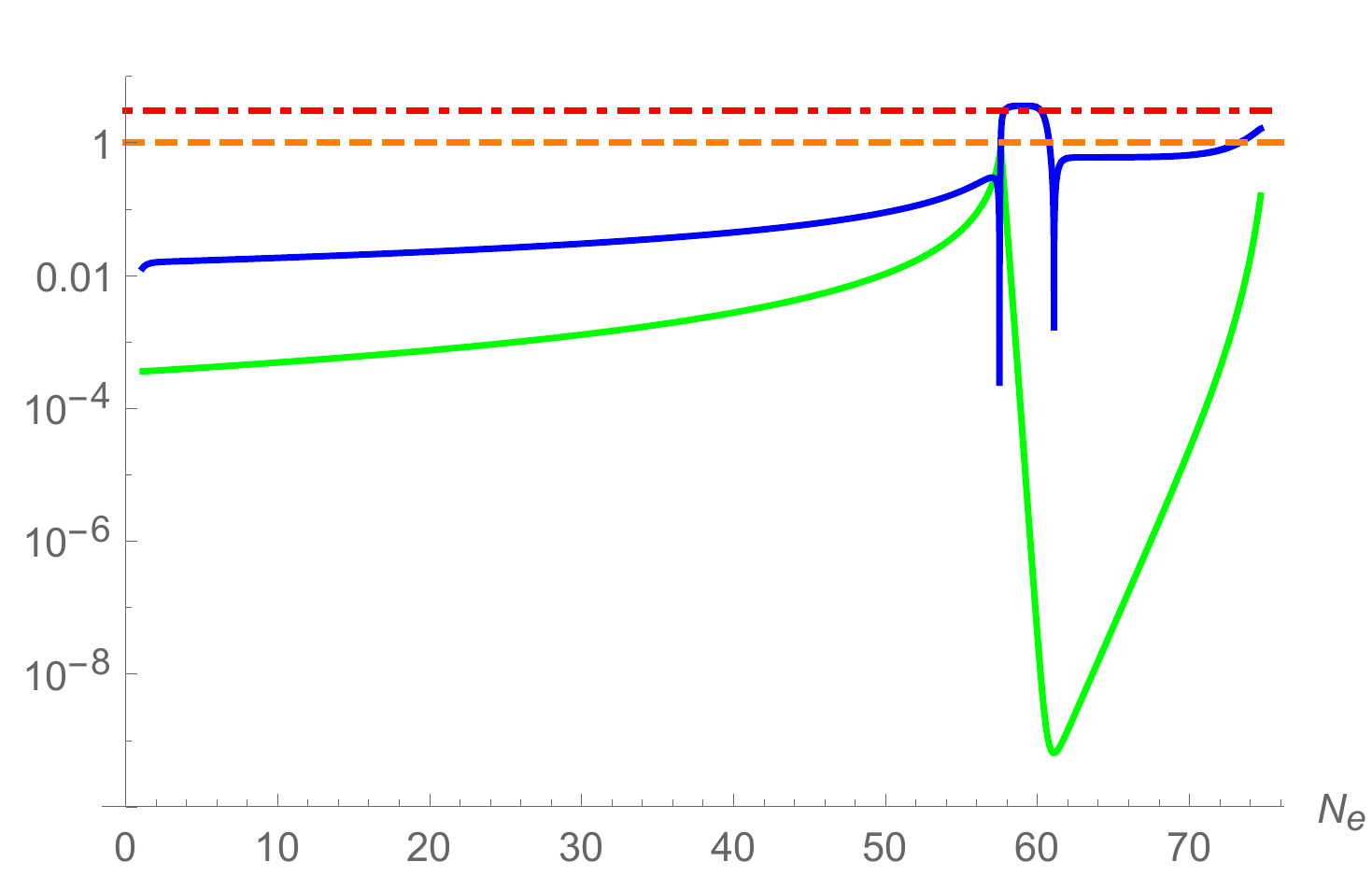}
    \raisebox{1.2\height}{\includegraphics[width=0.05\textwidth]{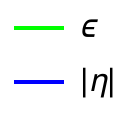}}
    
    \includegraphics[width=0.4\textwidth]{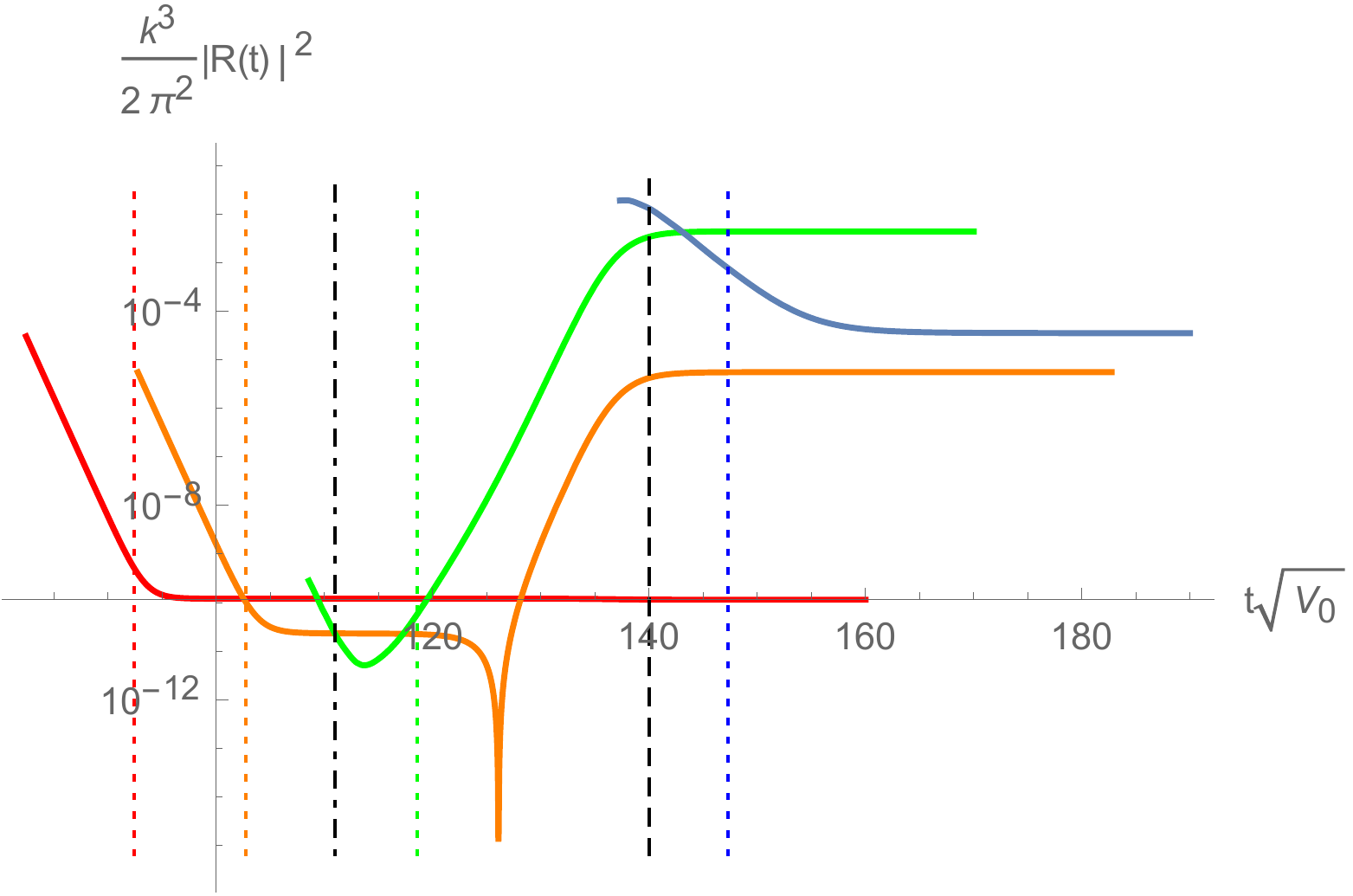}
    \includegraphics[width=0.4\textwidth]{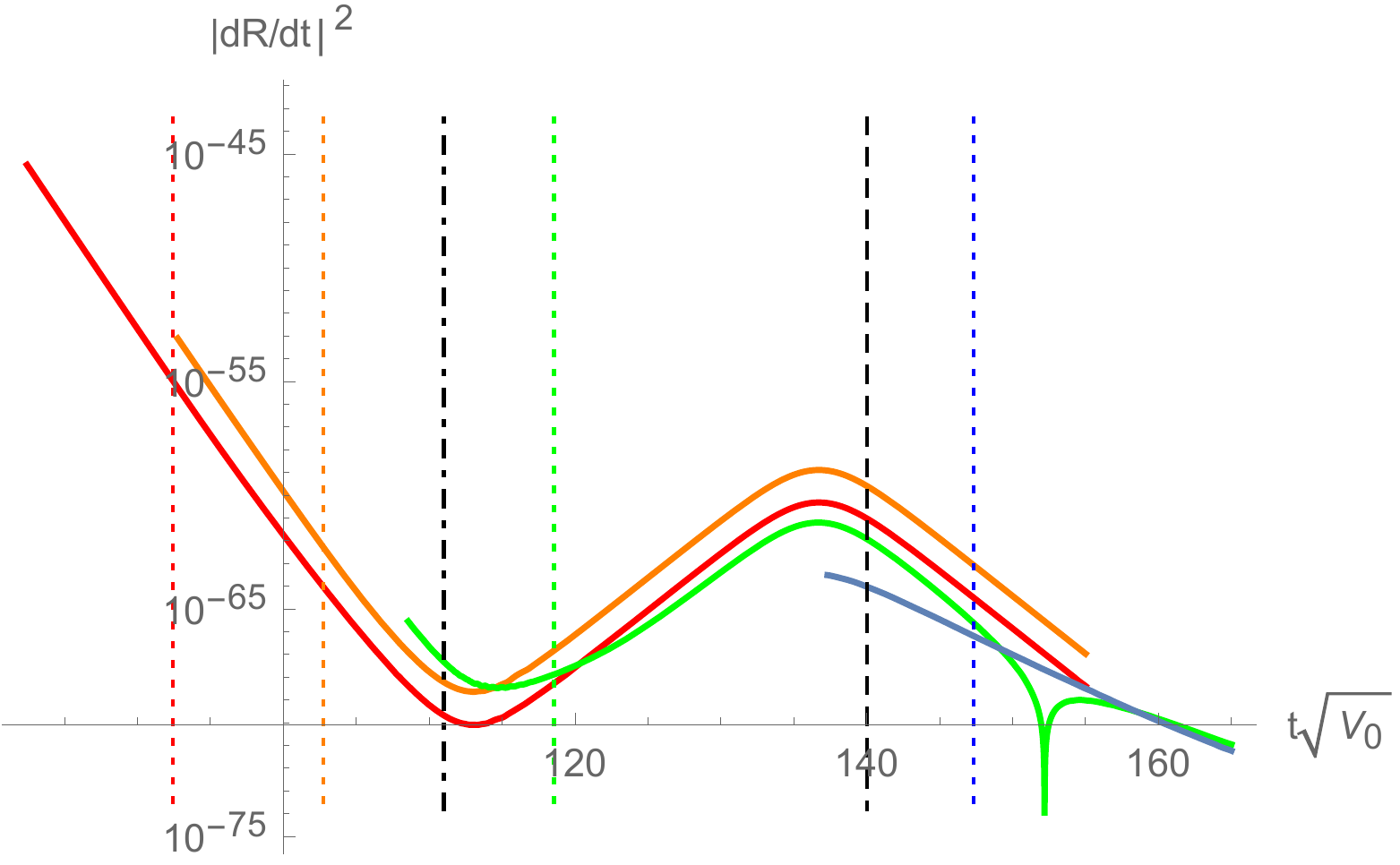}
    \raisebox{0.7\height}{\includegraphics[width=0.1\textwidth]{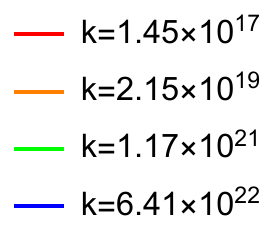}}
    \caption{In the upper panel, we plot the evolution of $\epsilon$ and $\eta$ as functions of rescaled time $t\sqrt{V_{0}}$~(left) and the e-folding numbers from the end of inflation $N_{e}$~(right). In the lower panel, we plot the time evolution of $\frac{k^{3}}{2\pi^{2}}|\mathcal{R}(t)|^{2}$ and $|\dot{R}|^{2}$ for $k<k_{min}$~(red line), $k_{min}<k<k_{s}$~(orange line), $k_{s}<k<k_{e}$~(green line), $k>k_{e}$~(blue line). The vertical dotted line denote the horizon crossing time for each mode with the same color. The gray dotdashed line and the gray dashed line denote $t_{s}$ and $t_{e}$, respectively. 
    }
    \label{fig:k}
\end{figure*}
For the modes which are deep inside the horizon, i.e. $k\gg \mathcal{H}$, $\frac{z''}{z}$ is much smaller than $k^{2}$, so the solution of $u_{k}$ is taken to be the Bunch-Davies type~\cite{Bunch:1978yq}
\begin{equation}
u_{k} = \frac{e^{-i k \tau}}{\sqrt{2 k}}.
\end{equation}

For the modes outside the horizon, i.e., $k\ll\mathcal{H}$, in the slow-roll inflationary models with $\epsilon,\eta\ll 1$, we can obtain from Eq.~\eqref{eq:tauRk}
\begin{equation}
\label{eq:dotdec}
\dot{\mathcal{R}}_{k}=A_{k} e^{-2H(t-t_{k})},
\end{equation}
where $A_{k}$ is constant and $t_{k}$ denotes the horizon crossing time~($k=a(t_{k})H$). In the next section, we will see that the assumption above is reasonable.
$\dot{\mathcal{R}}_{k}$ decreases exponentially, so $\mathcal{R}_{k}$ remains almost constant for the modes outside the horizon. 

In the USR regime, $\dot{\phi}$ is extremely small due to the flatness of $V(\phi)$, so $\epsilon$ is neglegible according to Eq.~\eqref{eq:srp}. Then for the modes with $k\ll \mathcal{H}$ in the USR regime, the EOM of $\mathcal{R}_{k}$ is approximately
\begin{equation}
\label{eq:oheom}
\ddot{\mathcal{R}}_{k}+(3-2\eta_{m})H\dot{\mathcal{R}}_{k}=0,
\end{equation}
which means $\dot{\mathcal{R}}_{k}$ evolves as
\begin{equation}
\label{eq:dotinc}
\dot{\mathcal{R}_{k}}=A_{k} e^{-2H(t_{s}-t_{k})} e^{(2\eta_{m}-3)H(t-t_{s})}.
\end{equation}
For $\eta_{m}>3/2$, $\dot{\mathcal{R}}_{k}$ increases outside the horizon which causes superhorizon evolution of scalar perturbations. For example, if $dV/d\phi=0$, the EOM of $\phi$ reads $\ddot{\phi}+3H\dot{\phi}=0$, $\eta_{m}=3$ and $\dot{\mathcal{R}}_{k}\propto e^{3Ht}$.

The power spectrum of scalar perturbations is defined by
\begin{equation}
\label{eq:PRdef}
\mathcal{P_{R}}(k)=\dfrac{k^{3}}{2\pi^{2}}\left|\mathcal{R}_{k}(t_{f})\right|^{2}=\dfrac{k^{3}}{2\pi^{2}}\left|\dfrac{u_{k}(t_{f})}{z(t_{f})}\right|^{2},
\end{equation}
where the subscript $f$ denotes the end of inflation. 
The spectral index of scalar perturbations is defined by
\begin{equation}
\label{eq:index}
n_{\mathcal{R}}(k)-1=\dfrac{d\ln \mathcal{P_{R}}(k)}{d\ln k}.
\end{equation}
To clarify the superhorizon evolution, we define a new quantity $\mathcal{P}_{\mathcal{R}}^{H}(k)\equiv\frac{k^{3}}{2\pi^{2}}\left|\frac{u_{k}(t_{k})}{z(t_{k})}\right|^{2}$. Following~\cite{Stewart:1993bc}, we have
\begin{equation}
\label{eq:approPR}
\mathcal{P}_{\mathcal{R}}^{H}(k)=2^{2\nu-3} \left(\frac{\Gamma(\nu)}{\Gamma\left(\frac{3}{2}\right)}\right)^{2}(1-\epsilon(t_{k}))^{2\nu-1}\dfrac{H^{2}}{8\pi^{2}\epsilon(t_{k})}.
\end{equation}
where $\nu=\frac{1-\eta+\epsilon}{1+\epsilon}+\frac{1}{2}$.

Under the slow-roll conditions, $\mathcal{R}_{k}$ remains constant at superhorizon scales, so $\mathcal{P_{R}}(k)=\mathcal{P}^{H}_{\mathcal{R}}(k)$. The scalar spectral index is
\begin{equation}
\label{eq:index1}
n_{\mathcal{R}}(k)-1=-4\epsilon+2\eta.
\end{equation} 
In the USR models with $\eta_{m}>3/2$,there are some modes $k$ with $\mathcal{P_{R}}(k)>\mathcal{P}^{H}_{\mathcal{R}}(k)$, according to Eq.~\eqref{eq:dotinc}.

For the modes with $k<k_{s}$, using~\eqref{eq:dotdec} and~\eqref{eq:dotinc}, one can obtain
\begin{equation}
\label{eq:Rkdot}
\begin{split}
\dot{\mathcal{R}}_{k}(t_{s})&=A_{k}e^{-2H(t_{s}-t_{k})},\\
\dot{\mathcal{R}}_{k}(t_{e})&=A_{k}e^{(2\eta_{m}-3)H(t_{e}-t_{s})-2H(t_{s}-t_{k})},\\
\end{split}
\end{equation}
which implies $\dot{\mathcal{R}}$ decreases from $t_{k}$ to $t_{s}$ and then increases from $t_{s}$ to $t_{e}$. The modes which satisfy $2(t_{s}-t_{k})>(2\eta_{m}-3)(t_{e}-t_{s})$ remain constant outside the horizon because $\dot{\mathcal{R}}_{k}$ is always much smaller than $\dot{\mathcal{R}}_{k}(t_{k})$. 
The condition, $2(t_{s}-t_{k})>(2\eta_{m}-3)(t_{e}-t_{s})$, can be transformed into
\begin{equation}
k<k_{s}e^{-H(2\eta_{m}-3)(t_{e}-t_{s})/2}=k_{min}.
\end{equation}

For the modes with $k_{min}<k<k_{s}$, according to Eq.~\eqref{eq:Rkdot}
, $\mathcal{R}_{k}$ is obtained as
\begin{equation}
\label{eq:Rkt}
\begin{split}
\mathcal{R}_{k}(t_{e})=&\dfrac{A_{k}}{2H}\left(e^{-2Ht_{k}}-e^{-2Ht_{s}}\right)+\mathcal{R}_{k}(t_{k})+\\
&\left(\dfrac{k}{k_{s}}\right)^{2}\dfrac{A_{k}}{(2\eta_{m}-3)H}\left(e^{(2\eta_{m}-3)H(t_{e}-t_{s})}-1\right).
\end{split} 
\end{equation}
Since $\dot{\mathcal{R}}_{k}$ is exponentialy amplified after the horizon crossing, the second term in the r.h.s. of Eq.~\eqref{eq:Rkt} can be neglected\footnote{For those $k$ which are slightly larger than $k_{min}$, $\mathcal{P_{R}}(k)$ can be orders of magnitude smaller than $\mathcal{P}^{H}_{\mathcal{R}}(k)$, as stated in the next section.}. Since $\mathcal{R}_{k}(t_{f})=\mathcal{R}_{k}(t_{e})$, according to Eqs.~\eqref{eq:PRdef} and~\eqref{eq:index1}, for $k_{min}<k<k_{s}$ one can obtain
\begin{equation}
\label{eq:ns4}
\begin{split}
\mathcal{P_{R}}(k)&=\left(\dfrac{k}{k_{s}}\right)^{4}\mathcal{P}^{H}_{\mathcal{R}}(k)e^{H(4\eta_{m}-6)(t_{e}-t_{s})},\\
n_{\mathcal{R}}(k)-1&=4-4\epsilon(t_{k})+2\eta(t_{k}).
\end{split}
\end{equation}
Since $t_{k}$ is in the slow-roll regime, one can neglect $\epsilon(t_{k})$ and $\eta(t_{k})$ and obtain $n_{\mathcal{R}}(k)-1=4$, which is in agreement with the result of Ref.~\cite{Byrnes:2018txb}.

For the modes with $k_{s}<k<k_{e}$, 
since $\epsilon\ll\eta_{m}$ in the USR regime, from Eq.~\eqref{eq:rela} one can obtain $\epsilon\propto e^{-2H\eta_{m} t}$ . After leaving the horizon, $\mathcal{R}_{k}$ continues to increase until $t_{e}$ as $\mathcal{R}_{k}\propto e^{(2\eta_{m}-3)Ht}$. After some calculation, one can obtain
\begin{equation}
\label{eq:PRkc}
\begin{split}
\mathcal{P_{R}}(k)&=\mathcal{P}^{H}_{\mathcal{R}}(k)e^{2(2\eta_{m}-3)H(t_{e}-t_{k})}\\
&=\left(\dfrac{k}{k_{s}}\right)^{6-2\eta_{m}}\mathcal{P}^{H}_{\mathcal{R}}(k_{s})e^{H(2(2\eta_{m}-3)t_{e}+(6-4\eta_{m}) t_{s})},\\
n_{\mathcal{R}}-1&=6-2\eta_{m}.
\end{split}
\end{equation} 

For the modes with $k>k_{e}$, after horizon crossing the slow-roll conditions are valid, so $\mathcal{P_{R}}(k)=\mathcal{P}^{H}_{\mathcal{R}}(k)$.

We consider the inflationary model in Ref.~\cite{Cicoli:2018asa} as an example to present the evolution of $\epsilon$, $\eta$, $\mathcal{R}_{k}$ and $\dot{\mathcal{R}}_{k}$ numerically. The effective potential reads
\begin{equation}
\label{eq:pot}
\begin{split}
V(\phi)=V_{0}&\left[1-\dfrac{e^{-\frac{1}{\sqrt{3}} \phi}}{C_{1}}\left(1-\frac{C_{2}}{1-C_{3} e^{-\frac{1}{\sqrt{3}} \phi}}\right) \right.\\
&\left. -  \frac{C_{4}e^{\frac{2}{\sqrt{3}}\phi}}{C_{1}(1+C_{5} e^{\sqrt{3} \phi})}\right].
\end{split}
\end{equation}
The parameters are chosen as $V_{0}=3.5\times 10^{-10}$, $C_{1}=0.360335$, $C_{2}=0.5$, $C_{3}=0.264443$, $C_{4}=4.16459\times 10^{-2}$ and $C_{5}=3.82375\times 10^{-2}$. 

In the upper panel of Fig.~\ref{fig:k}, we show the time evolution of $\epsilon$ and $\eta$. We can see that $\eta$ quickly changes around $t_{s}$ and $t_{e}$ and remains nearly constant during the USR regime. $\epsilon$ exponentially decreases between $t_{s}$ and $t_{e}$. 
In the lower panel of Fig.~\ref{fig:k}, we show the time evolution of $\mathcal{R}_{k}$ and $\dot{\mathcal{R}}_{k}$ for $k<k_{min}$~(red line), $k_{min}<k<k_{s}$~(orange line), $k_{s}<k<k_{e}$~(green line) and $k>k_{e}$~(blue line). The horizon crossing time for each mode is depicted by vertical dotted lines with corresponding colors. $\dot{\mathcal{R}}_{k}$ exponentially increases in the USR regime, and for $k_{min}<k<k_{e}$, $\mathcal{R}_{k}$ increases at superhorizon scales.

\section{Analytical result of $\mathcal{P_{R}}(k)$}
\label{sec:PR}

In this section, we present the analytical method of calculating $\mathcal{P_{R}}(k)$ in the USR models, and find the expression of the spectral index of $\mathcal{P_{R}}(k)$ in terms of the inflationary potential. Then, we compare the numerical result with the analytical result of $\mathcal{P_{R}}(k)$ in the model with~\eqref{eq:pot}. 


For arbitrary USR models, one can expand the potential near the USR regime. According to Eq.~\eqref{eq:eomphi}, due to the smallness of $|\dot{\phi}|$, $\phi$ changes very slowly in the USR regime, so $\phi-\phi(t_{e})$ is a small quantity. Let $\phi_{e}\equiv\phi(t_{e})$, the Taylor expansion of $V(\phi)$ at $\phi_{e}$ reads
\begin{equation}
\label{eq:Vphi}
V(\phi)=b_{0}+b_{1}(\phi-\phi_{e})+b_{2}(\phi-\phi_{e})^{2}+\cdots,
\end{equation}
where the higher-order terms in Eq.~\eqref{eq:Vphi} can be neglected. According to Eq.~\eqref{eq:srp}, $|\dot{\phi}|$ also reaches the minimum at $t_{e}$, so $\ddot{\phi}(t_{e})=0$ and $3H\dot{\phi}(t_{e})=-\frac{dV}{d\phi}(t_{e})=-b_{1}$. To guarantee $|\dot{\phi}|$ increases after $t_{e}$, $b_{2}$ should be negative. Since the term $\dot{\phi}^{2}/2$ is negligible, $H$ can be estimated as $\sqrt{b_{0}/3}$. 
Then, one can solve the EOM of $\phi$ and obtain
\begin{equation}
\label{eq:phievo0}
\begin{split}
&\phi-\phi_{e}+\dfrac{b_{1}}{2b_{2}}=\dfrac{1}{4b_{2}\alpha}\times\\
&\left[\left(-3Hb_{1}+\alpha b_{1}+\frac{4b_{2}}{3H}b_{1}\right)e^{(-3H-\alpha)(t-t_{e})/2}+\right.\\
&\left.\left(3Hb_{1}+\alpha b_{1}-\frac{4b_{2}}{3H}b_{1}\right)e^{(-3H+\alpha)(t-t_{e})/2}\right],
\end{split}
\end{equation}
where $\alpha=\sqrt{3b_{0}-8b_{2}}$.
\begin{figure}[h]
    \includegraphics[width=0.4\textwidth]{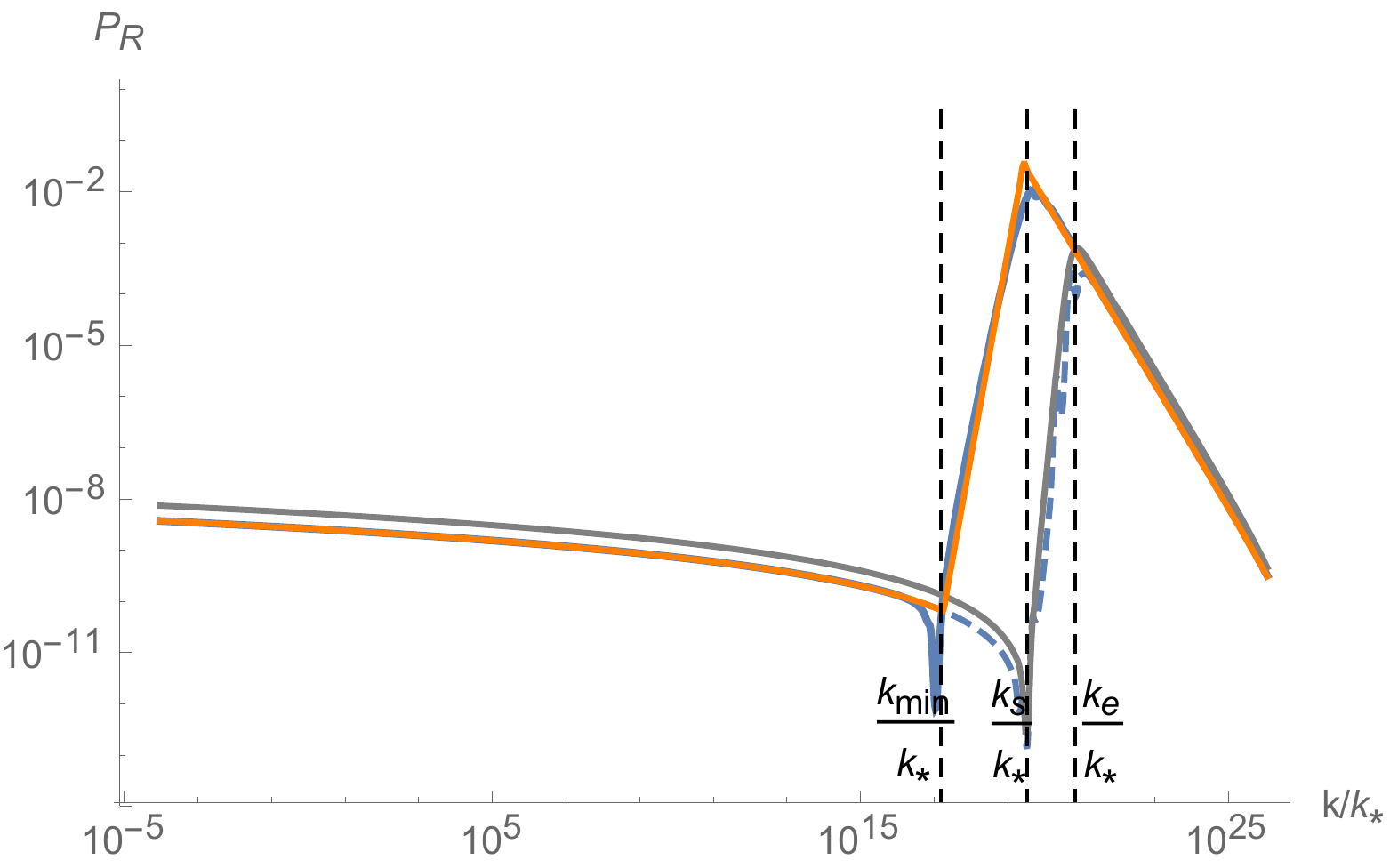}
    
    \caption{The numerical result~(blue solid line) and analytical result~(orange line) of $\mathcal{P_{R}}(k)$, the approximate result with~\eqref{eq:approPR}~(blue dashed line) and numercal result of $\mathcal{P}^{H}_{\mathcal{R}}(k)$, where $k_{*}=0.05\mathrm{Mpc}^{-1}$ in the horizontal axis.
    The numerical result of $\mathcal{P_{R}}(k)$ is apparently larger than $\mathcal{P}^{H}_{\mathcal{R}}$~(gray line) for $k_{min}<k<k_{e}$ which manifests the superhorizon evolution. The verticle dashed lines from left to right denote $k_{min}/k_{*}$, $k_{min}/k_{*}$ and $k_{min}/k_{*}$, respectively.
    }
    \label{fig:PR}
\end{figure}
The first and second terms in the r.h.s. of Eq.~\eqref{eq:phievo0} dominate before $t_{e}$ and after $t_{e}$, respectively. Due to the exponential dependence on $t$, $\eta$ changes from $\frac{3H+\alpha}{2H}$ to $\frac{-3H+\alpha}{2H}$ quickly around $t_{e}$. According to Eq.~\eqref{eq:PRkc} and Eq.~\eqref{eq:index1}, the spectral index for both $k_{s}<k<k_{e}$ and $k>k_{e}$ is\footnote{
    In the case of $b_{2}=0$, $V(\phi)$ is extremely flat near $\phi_{e}$, which causes that $|n_{\mathcal{R}}-1|$ is close to zero.}
\begin{equation}
\label{eq:n2}
n_{\mathcal{R}}-1=3-\frac{\alpha}{H}.
\end{equation}

For the model with~\eqref{eq:pot}, one can obtain $\phi_{e}=1.2726199393$, $b_{1}=7.48205\times 10^{-16}$, $b_{2}=-7.30823\times 10^{-12}$.
In Fig.~\ref{fig:PR}, we show the analytical result~(orange line) and the numerical result~(blue solid line) of $\mathcal{P_{R}}(k)$. The approximate result from~\eqref{eq:approPR}~(blue dashed line) and numerical result~(gray line) of $\mathcal{P}^{H}_{\mathcal{R}}(k)$ are also shown as comparison, which indicates that scalar perturbations evolve outside the horizon and the approximate result~\eqref{eq:approPR} is not valid in the USR models. Because of the assumption that before $t_{s}$ the slow-roll conditions are valid, we neglect $\epsilon$ and $\eta$ in Eq.~\eqref{eq:ns4}, simply set $n_{\mathcal{R}}(k)-1=4$ for the modes with $k_{min}<k<k_{s}$ in the analytical result. As shown in Fig.~\ref{fig:k}, around $t_{s}$, $\epsilon$ is no longer a small quantity and the decrease of $\mathcal{P}^{H}_{\mathcal{R}}$ cannot be neglected. This is the reason that around $k_{s}$ the analytical result is slightly larger than the numerical result. For the modes with $k>k_{s}$, from Eq.~\eqref{eq:n2} we have $n_{\mathcal{R}}(k)-1=-1.198$ which agrees well with the numerical result. Both the analytical and numerical result indicate that $\mathcal{P_{R}}$ peaks at $k_{s}$.


From Eq.~\eqref{eq:oheom} we can obtain the sign of $\mathcal{R}_{k}(t_{k})$ and $A_{k}$ are different, so the second and third terms in the r.h.s. of Eq.~\eqref{eq:Rkt} also have different signs. The second term is dominant for $k<k_{min}$ while the third term is dominant for $k>k_{min}$, so there exists such a mode close to $k_{min}$ where the two terms cancel each other. This causes the sudden decrease of $\mathcal{P_{R}}$ near $k_{min}$ in the numerical result, as shown in Fig.~\ref{fig:PR}. See Ref.~\cite{Ozsoy:2019lyy} for detailed discusstion of this phenomenon.


\section{induced SGWB}
\label{sec:GW}

It is well-known that tensor perturbations are coupled to scalar perturbations at the second order, and the amplified scalar perturbations may cause a scalar-induced SGWB which may be detected by GW experiments. In this section, we briefly review the methods to calculate the energy spectrum $\Omega_{\mathrm{GW}}$ of the induced GWs derived in Ref.~\cite{Kohri:2018awv}, and find the relationship between the spectral index of $\Omega_{\mathrm{GW}}$ and $V(\phi)$.

Recall the perturbed metric with~\eqref{eq:meper}, the Fourier modes of $h_{ij}$ are introduced as
\begin{equation}
h_{i j}(\tau, \mathbf{x})=\int \frac{d^{3} \mathbf{k}}{(2 \pi)^{3 / 2}} e^{i \mathbf{k} \cdot \mathbf{x}}\left[h_{\mathbf{k}}^{+}(\tau) \mathrm{e}_{i j}^{+}(\mathbf{k})+h_{\mathbf{k}}^{\times}(\tau) \mathrm{e}_{i j}^{\times}(\mathbf{k})\right],
\end{equation}
where $\mathrm{e}_{i j}^{+}(\mathrm{k})$ and $\mathrm{e}_{i j}^{\times}(\mathrm{k})$ are polarization tensors which satisfy $\sum_{i, j} \mathrm{e}_{i j}^{\alpha}(\mathrm{k}) \mathrm{e}_{i j}^{\beta}(-\mathrm{k})=\delta^{\alpha \beta}$.
The EOM of $h_{\mathbf{k}}$ is obtained from the Enstein equation to the second order
\begin{equation}
h_{\mathbf{k}}^{\prime \prime}+2 \mathcal{H} h_{\mathbf{k}}^{\prime}+k^{2} h_{\mathbf{k}}=S(\tau, \mathbf{k}),
\end{equation}
where $S(\tau, \mathbf{k})$ is the Fourier transformation of the source term $S_{i j}(\tau, \mathbf{x})$,
\begin{equation}
\label{eq:tensoreomf}
S(\tau, \mathbf{k}) = -4 {\mathrm e}^{i j}(\mathbf{k}) \int \frac{d^3 \mathbf{x}}{(2 \pi)^{3 / 2}} e^{-i \mathbf{k} \cdot \mathbf{x}} S_{i j}(\tau, \mathbf{x}),
\end{equation}
and $S(\tau, \mathbf{k})$ is defined by~\cite{Baumann:2007zm,Ananda:2006af}
\begin{equation}
\begin{split}
S_{i j}(\tau, \mathbf{x})=&4 \Phi \partial_{i} \partial_{j} \Phi+2 \partial_{i} \Phi \partial_{j} \Phi\\
&-\frac{4}{3(1+w) \mathcal{H}^{2}} \partial_{i}\left(\Phi^{\prime}+\mathcal{H} \Phi\right) \partial_{j}\left(\Phi^{\prime}+\mathcal{H} \Phi\right).
\end{split}
\end{equation}
The energy spectrum $\Omega_{\mathrm{GW}}$ of the SGWB is defined by
\begin{equation}
\Omega_{\mathrm{GW}}(\tau, k)=\frac{1}{24}\left(\frac{k}{\mathcal{H}(\tau)}\right)^{2} \overline{\mathcal{P}_{h}(\tau, k)}
\end{equation}
where the two polarization modes of GWs have been summed over and $\mathcal{P}_{h}$ is the tensor perturbation spectrum. The overline denotes the average among several wavelengths.

\begin{figure}[h]
    \includegraphics[width=0.4\textwidth]{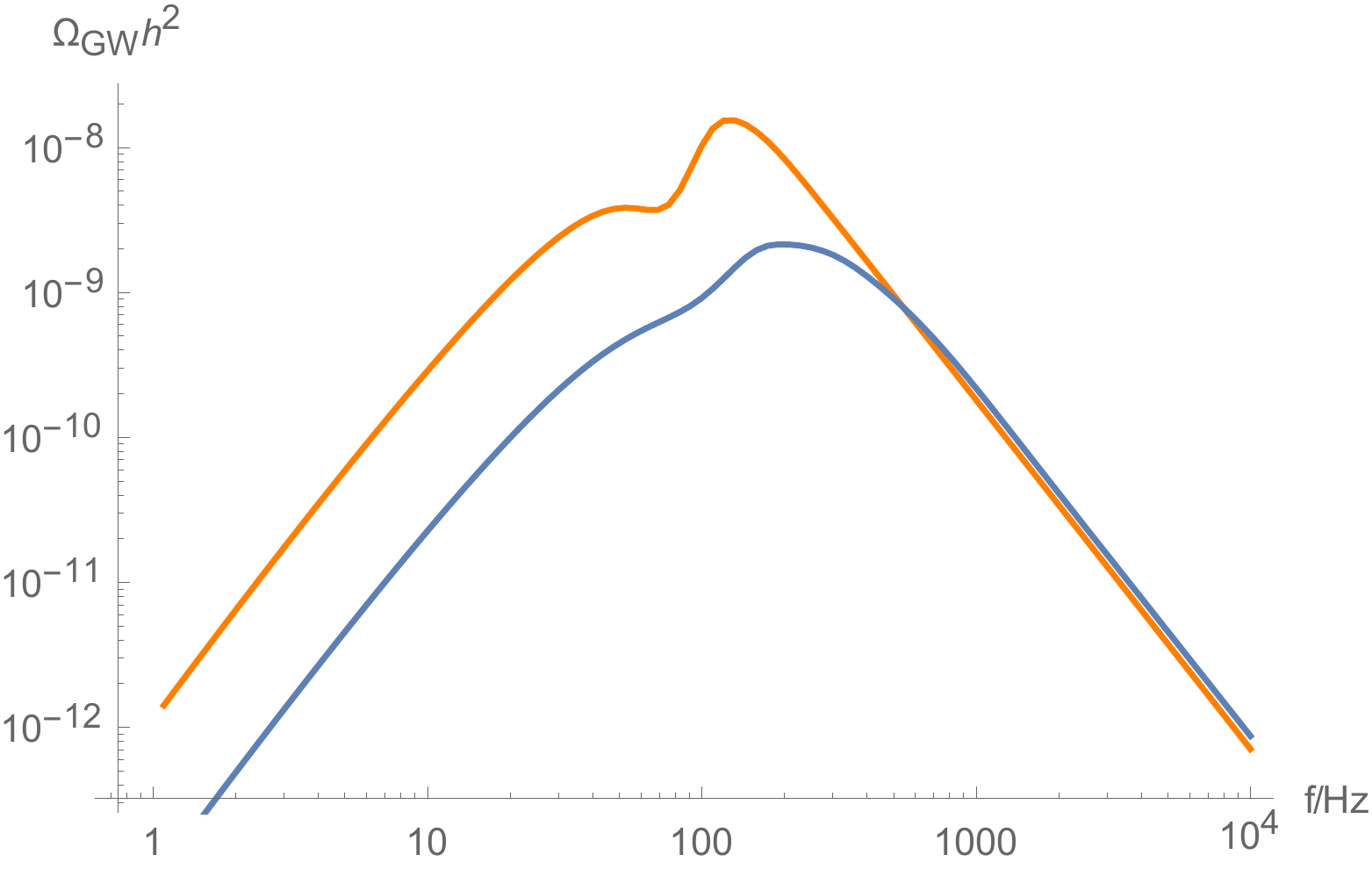}
    \caption{The numerical result of $\Omega_{\mathrm{GW}}$ calculated from analytical approximate result~(orange line) and numerical result~(blue line) of $\mathcal{P_{R}}$.
    }
    \label{fig:GW}
\end{figure}

In Fig.~\ref{fig:GW}. we show the induced SGWB using the analytical~(orange line) and numerical result~(blue line) of $\mathcal{P_{R}}$ obtained in Sec.~\ref{sec:PR}. 
Applying the analytical results of Ref.~\cite{Xu:2019bdp}, $\Omega_{\mathrm{GW}}(k)$ also peaks at $k_{s}$, and 
\begin{equation}
 \Omega_{\mathrm{GW}}(k)\propto\left\{
     \begin{aligned}
     &k^{3},& {\rm for} \ k<k_{s},\\
     &k^{6-2\alpha/H},&{\rm for} \ k>k_{s},
     \end{aligned}
     \right.
\end{equation}
which coincides with the numerical result. Because of the deviation between analytical and numerical result of $\mathcal{P_{R}}$ around $k_{s}$, the peak value of two curves in Fig.~\ref{fig:GW} differs by about one order of magnitude, while the spectral indexes coincide with each other for $k\gg k_{s}$ and $k\ll k_{s}$.



Let $n_{IR}$ and $n_{UV}$ denote the spectral indexes of $\Omega_{\mathrm{GW}}$ in the infrared regime and the ultraviolet regime, respectively. $n_{IR}=3$ is a universal result as claimed in Ref.~\cite{Cai:2019cdl}, while $n_{UV}$ is in general different for different SGWBs. As reported in Ref.~\cite{Kuroyanagi:2018csn}, $n_{UV}$ is fixed for the SGWB sourced by domain walls~($n_{UV}=-1$), cosmic strings~($n_{UV}=0$), bubble colision~($n_{UV}=-2$), turbulence~($n_{UV}=-5/3$), sound waves~($n_{UV}=-4$), kination models~($n_{UV}=1$) and self-ordering scalar fields~($n_{UV}=0$), while the SGWB from preheating are cutoff in the ultraviolet regime. In our case, $n_{UV}$ is proportional to a model parameter $b_{2}$. In the future, the detection of such a background with $n_{UV}$ different from those values of $n_{UV}$ above, it is likely to be generated from USR inflationary models, and from $\Omega_{\mathrm{GW}}$ one can conveniently extract the information of $V(\phi)$ the around USR regime.

\section{Conclusion and Discussion}
\label{sec:con}
In this work, we have presented an analytical method to estimate the scalar power spectrum in the USR inflationary models where numerical simulations are required and the usual 
slow-roll approximation breaks.  Assuming the transition between the USR regime and the slow-roll regime is instantaneous, we find $\mathcal{P_{R}}$ peaks at the Hubble-horizon scale when the USR process starts. On both sides of the peak, the USR process gives rise to a power-law behavior of frequency in $\mathcal{P_{R}}$ and the spectral indexes can be expressed in terms of the inflationary potential around the USR regime. 

Benefiting from the relationship between the spectral indexes of $\Omega_{\mathrm{GW}}$ and $\mathcal{P_{R}}$ discussed in Ref.~\cite{Xu:2019bdp}, one can reconstruct the inflationary potential near the USR regime from $\Omega_{\mathrm{GW}}$, and determine the spectral index of $\Omega_{\mathrm{GW}}$ from the inflationary potential.

\begin{acknowledgements}
    This work is supported in part by the National Natural Science Foundation of China Grants
    No.11690021, No.11690022, No.11851302, No.11947302 and No.11821505,
    in part by the Strategic Priority Research Program of the Chinese Academy of Sciences Grant No. XDB23030100
    and by Key Research Program of Frontier Sciences, CAS.
\end{acknowledgements}

\bibliography{ana}

\end{document}